\documentclass{ws-ijmpa}

\begin{document}

\markboth{Adam L. Scott}
{Search for Long-Lived Parents of the $Z^0$ Boson}

%
\catchline{}{}{}{}{}
%

\title{Search for Long-Lived Parents of the $Z^0$ Boson}

\author{\footnotesize Adam L. Scott\footnote{Representing the CDF Collaboration}
}

\address{Physics Department, University of California at Santa Barbara\\
Santa Barbara, California, 93106-9530,
United States of America}

%

\maketitle

\pub{Received 7 October 2004}{}

\begin{abstract}

We present the results of a search for new particles with long lifetime that decay to a $Z^0$ boson.  A long-lived parent of
the $Z^0$ is predicted by several models in addition to being an experimentally clean channel.  We vertex dimuons with invariant mass near
the $Z^0$ peak and study the decay length distribution.  No evidence of a long-lived component is found, and cross-section
limits are presented on a fourth generation quark model.

\keywords{Z boson; fourth generation quark; SUSY; displaced vertex.}
\end{abstract}

\section{\label{sec:introduction}Introduction}

We present a search for long-lived particles decaying to $Z^0$ bosons in $p\bar{p}$ collisions at $\sqrt{s}=1.96$~TeV with
the CDF detector\cite{bib:cdfdet} at the Tevatron.  We search for non-prompt sources of $Z^0\rightarrow \mu^+\mu^-$ by
vertexing muons and searching for an excess at large transverse distances from the beam, $L_{xy}$.  These results extend
previous searches\cite{bib:run1}.

There are several models that predict a long-lived particle decaying to a $Z^0$.  For example, gauge-mediated SUSY models
where the gravitino is the LSP\cite{bib:gravlsp} allow a long lifetime.  In a 4$^{th}$ generation quark model, if
$m_{b^\prime} < m_{t}$, its decay is $b^\prime \rightarrow b Z^0$ via a loop diagram, causing a long lifetime.
Beyond model dependent motivations, this is an experimentally appealing channel limited only by
track reconstruction.

\section{\label{sec:data}Data Sample \& Event Selection}

This search is performed using 163~pb$^{-1}$ of data collected between March 2002 and September 2003.  
The trigger requires a muon with $p_T > 18$~GeV/c.  
We select $Z^0$ events offline having two
muons with $p_T > 20$~GeV/c, well-measured tracks, and an invariant mass within 10 GeV of the $Z^0$ peak.  When the two
tracks are nearly back-to-back in $\phi$, a small offset in the impact parameter of one track can lead to a large shift in
$L_{xy}$, so we require $\Delta \phi < 175^{\circ}$, rejecting 99.8\% of background and remaining efficient for signal.
High transverse momentum of the $Z^0$ is a generic feature of signals, so we require $p_T^Z > 30\ \mathrm{GeV}$.
We search both with and without this cut, broadening the model independence and adding sensitivity at long
lifetimes.

We define, {\it a priori},  a minimum $L_{xy}$ for the signal region based on the expected background calculated from
Standard Model $Z^0$ generated with \texttt{PYTHIA}\cite{bib:pythia} and processed with a full detector simulation.  The
requirement is $L_{xy}>0.3~{\rm mm}$, which is tightened to $L_{xy}>1.0~{\rm mm}$ in the case without a $p_T^Z$ cut.

\section{\label{sec:backgrounds}Backgrounds}

The dominant background is from Standard Model $Z^0$ bosons where mis-reconstruction of one of the muon tracks produces a
large $L_{xy}$.  We measure this background from Monte Carlo using the data in the non-signal region to tune the Monte
Carlo resolution.  Three tuning methods are used and the differences are taken as a systematic uncertainty on the
background prediction.  We find backgrounds of $1.1 \pm 0.8$ events in the $p_T^Z>30\ \mathrm{GeV}$ case and $0.72 \pm 0.27$
events without the $p_T^Z$ cut.

Another background is from QCD events with large $L_{xy}$.  We estimate this background using Monte Carlo normalized to the
number of large $L_{xy}$ events in the data in an independent mass window, and find a background of $0.06 \pm 0.06$ events.
Cosmic rays, our final background, have inherently large $L_{xy}$.  We estimate this background from the number of events
removed by cosmic rejection cuts together with the efficiencies of those cuts and find a background of $0.0004 \pm 0.0001$
events.

\section{\label{sec:ae}Acceptance $\times$ Efficiency}


We use $b'$ Monte Carlo to quantify the acceptance$\times$efficiency as a function of lifetime and mass, and normalize the
efficiencies to those measured with Standard Model $Z^0$ candidates in data.  Uncertainties on the efficiencies arise from
the statistical precision with which they can be measured in the data and from systematic uncertainties in the simulation
modeling.  We find the dominant systematic uncertainty on the efficiency to be 7.4\% from simulation modeling.  Systematics
on acceptance arise from incomplete knowledge of initial and final state radiation (ISR/FSR) and parton distribution
functions (PDF).  We find a systematic uncertainty of 11.4\% on the acceptance.


\section{\label{sec:results}Results}

\begin{figure}[h]
\begin{center}
\mbox{
    \includegraphics[width=2.5in]{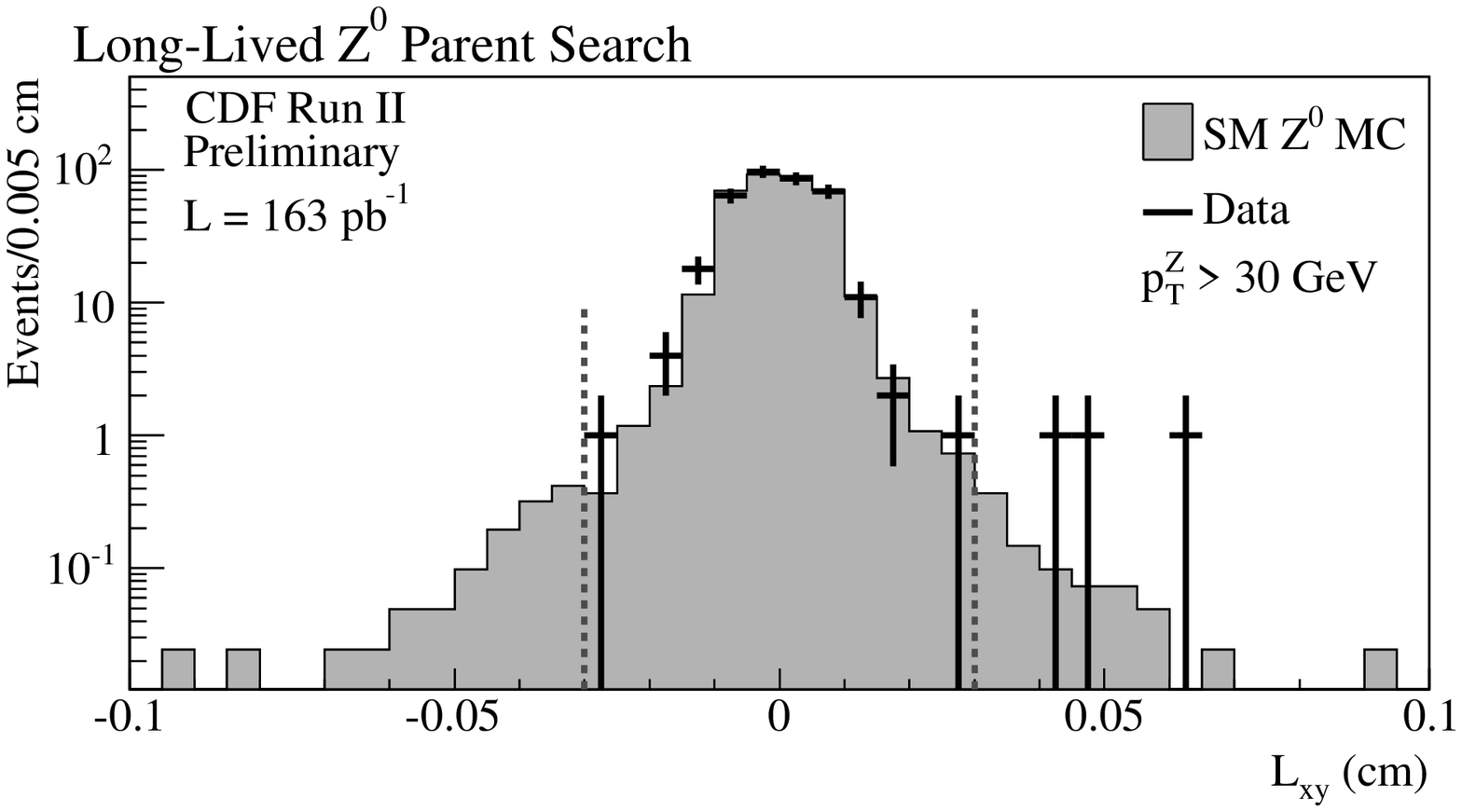}
    \includegraphics[width=2.5in]{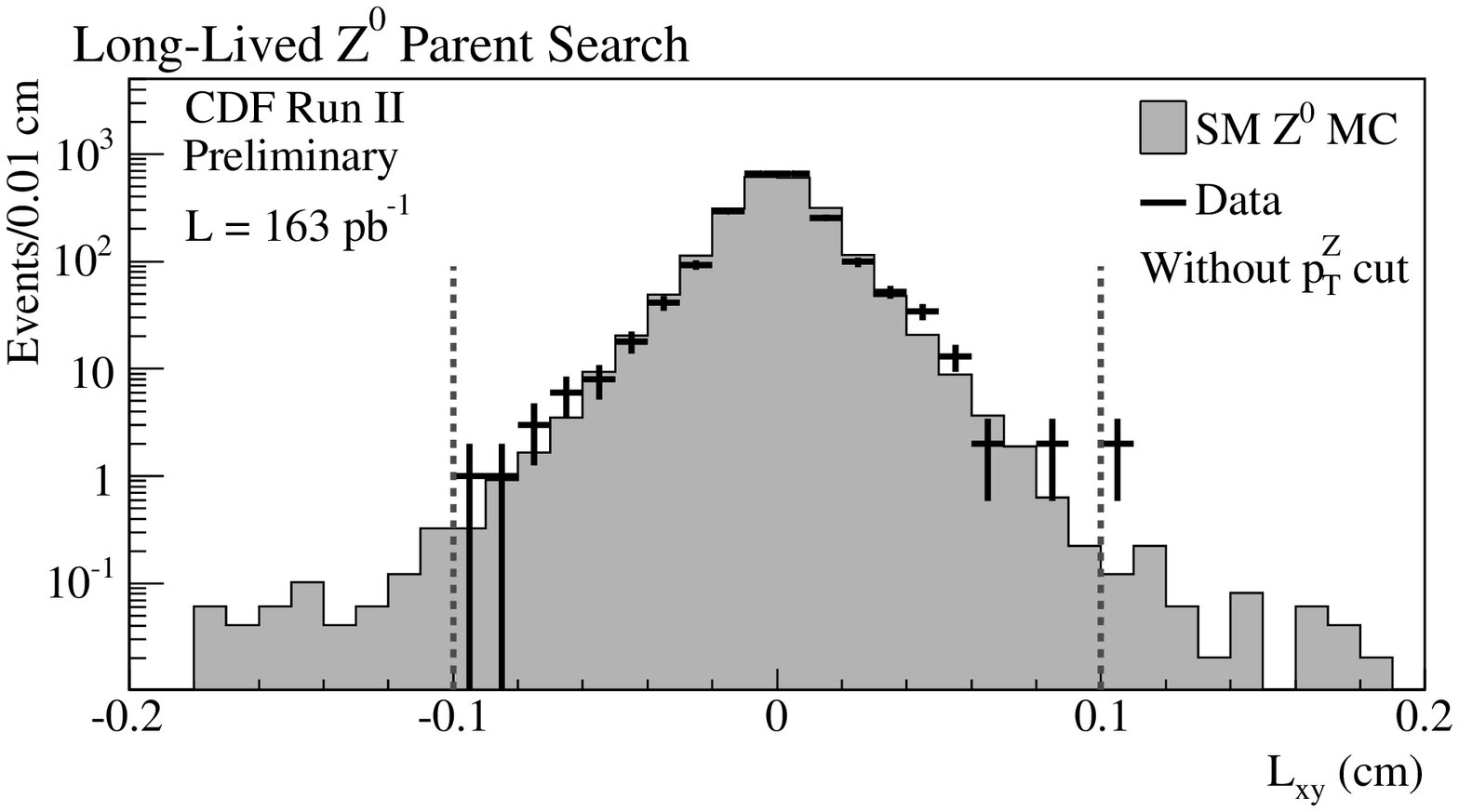}
}
\caption{$L_{xy}$ distributions in data (points) and Monte Carlo (histogram).
Left: With $p_T^Z > 30~\mathrm{GeV}$. Right: Without.
The dashed lines indicate the signal and control regions.}
\label{figures:data}
\end{center}
\end{figure}

We plot the $L_{xy}$ distribution in Figure~\ref{figures:data}.  We observe three events with $L_{xy} > 0.3\ \mathrm{mm}$
when we require $p_T^Z > 30\ \mathrm{GeV}$ 
and two events with $L_{xy} >
1.0\ \mathrm{mm}$ without the $p_T^Z$ cut.  In both cases, we see no events in the negative $L_{xy}$ control region.

\begin{figure}[h]
\begin{center}
\mbox{
    \includegraphics[width=2.5in]{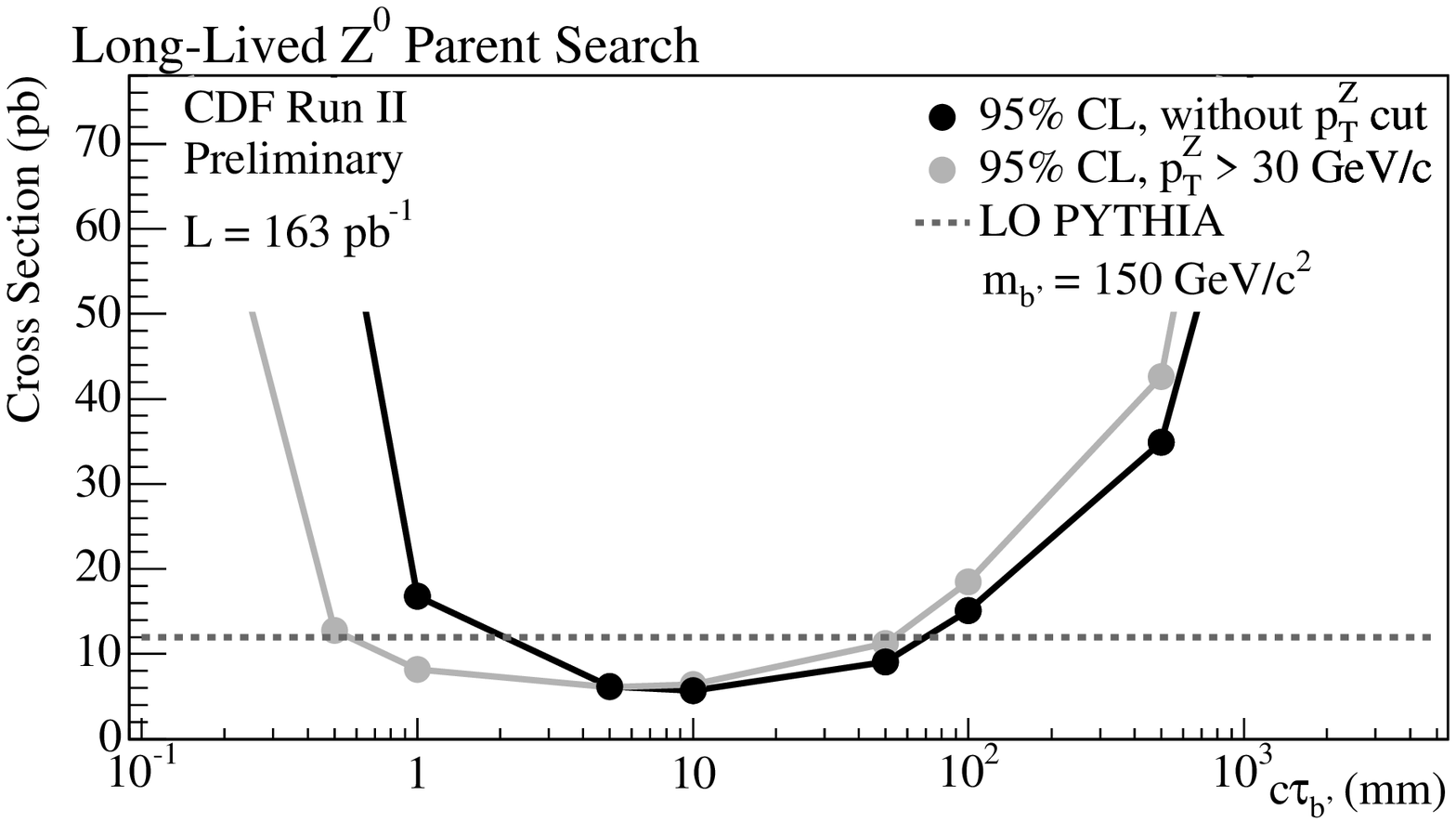}
    \includegraphics[width=2.5in]{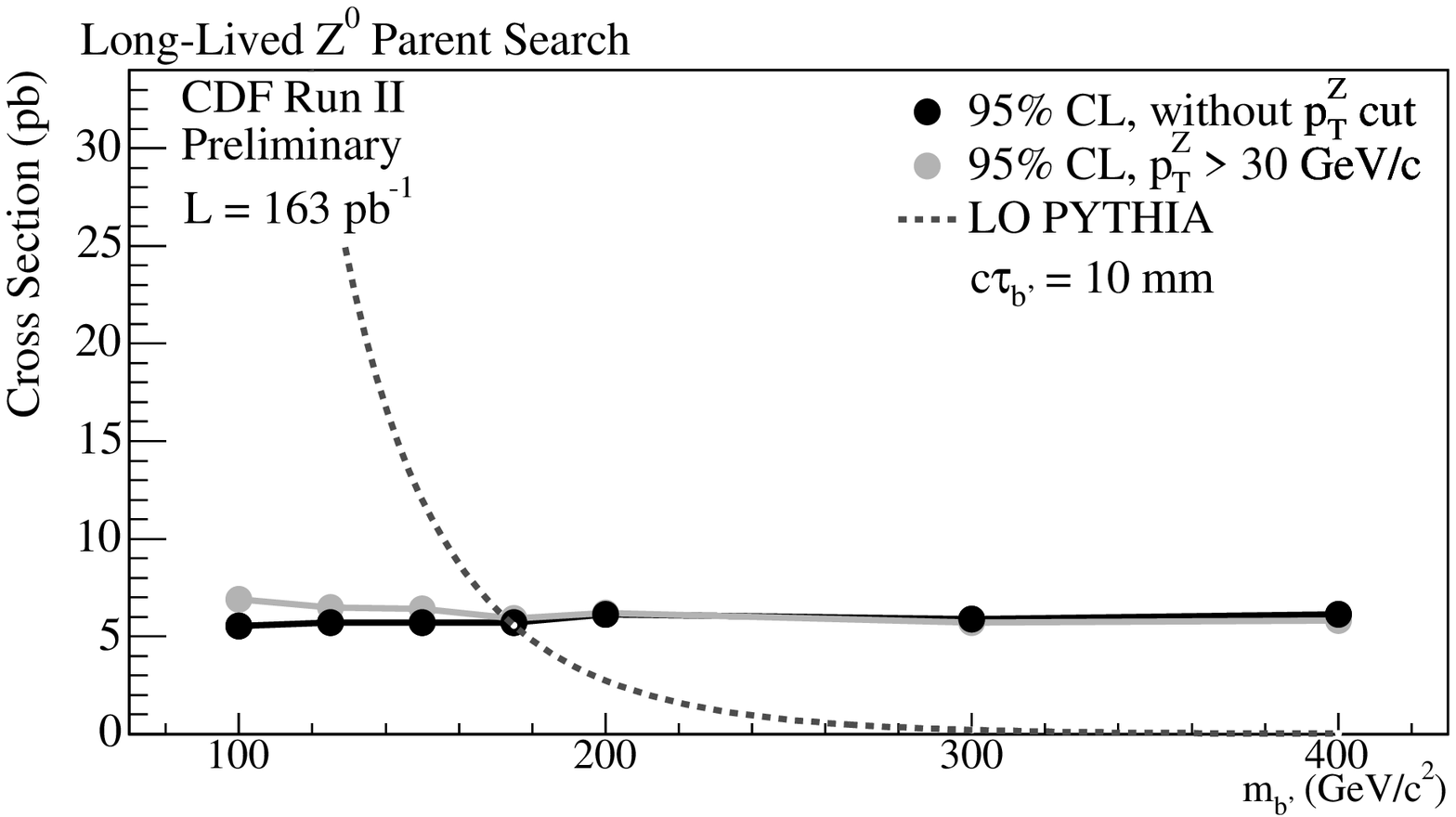}
}
\caption{Left: Cross section limits as a function of lifetime for
$m_{b'} = 150~\mathrm{GeV}/c^2$.
Right: Cross section limits as a function of mass for a lifetime of
$c\tau_{b'} = 10~{\rm mm}$.
}
\label{figures:fig_sigma}
\end{center}
\end{figure}

The observed number of events is consistent with the background expectation, and \textit{a posteriori} inspection of the
events shows them to be consistent with mis-reconstruction as expected for background.  
We calculate limits using the $b^\prime$ model for the acceptance$\times$efficiency.
The cross section limits are plotted as a function of lifetime and
mass in Figure~\ref{figures:fig_sigma} along with the LO theoretical cross section.


\section*{Acknowledgements}
\small{We thank the Fermilab staff and the technical staffs of the
participating institutions for their vital contributions. This
work was supported by the U.S. Department of Energy and National
Science Foundation; the Italian Istituto Nazionale di Fisica
Nucleare; the Ministry of Education, Culture, Sports, Science and
Technology of Japan; the Natural Sciences and Engineering Research
Council of Canada; the National Science Council of the Republic of
China; the Swiss National Science Foundation; the A.P. Sloan
Foundation; the Bundesministerium fuer Bildung und Forschung,
Germany; the Korean Science and Engineering Foundation and the
Korean Research Foundation; the Particle Physics and Astronomy
Research Council and the Royal Society, UK; the Russian Foundation
for Basic Research; the Comision Interministerial de Ciencia y
Tecnologia, Spain; and in part by the European 
Community's Human
Potential Programme under contract HPRN-CT-20002, Probe 
for New Physics.}

\end{document}